\documentclass[runningheads,a4paper]{llncs}

\usepackage{amsmath,amsfonts,amssymb,srcltx}
\usepackage{graphicx}

\usepackage{url}

\begin{document}

\mainmatter  

\title{A Computational Method for the Rate Estimation of Evolutionary Transpositions}

\titlerunning{Rate Estimation of Evolutionary Transpositions}

%
%
%
\author{Nikita Alexeev\inst{1,} \inst{2} \and Rustem Aidagulov\inst{3} \and Max A. Alekseyev\inst{1,}\thanks{Corresponding author. Email: \texttt{maxal@gwu.edu}}}
\authorrunning{N. Alexeev \and R. Aidagulov \and M.A. Alekseyev}
\institute{Computational Biology Institute, George Washington University, Ashburn, VA, USA\and
Chebyshev Laboratory, St. Petersburg State University, St. Petersburg,  Russia 
\and Department of Mechanics and Mathematics, Moscow State University, Moscow, Russia} 

%
%

\maketitle

\begin{abstract}
Genome rearrangements are evolutionary events that shuffle genomic architectures. 
Most frequent genome rearrangements are reversals, translocations, fusions, and fissions.
While there are some more complex genome rearrangements such as transpositions, they are rarely observed and 
believed to constitute only a small fraction of genome rearrangements happening in the course of evolution. 
The analysis of transpositions is further obfuscated by intractability of the underlying computational problems.

We propose a computational method for estimating the rate of transpositions in evolutionary scenarios between genomes.
We applied our method to a set of mammalian genomes and estimated the transpositions rate in mammalian evolution to be around 0.26.
\end{abstract}

\section{Introduction}

Genome rearrangements are evolutionary events that shuffle genomic architectures. 
Most frequent genome rearrangements are \emph{reversals} (that flip segments of a chromosome), \emph{translocations} (that exchange segments of two chromosomes),
\emph{fusions} (that merge two chromosomes into one), and \emph{fissions} (that split a single chromosome into two).
The minimal number of such events between two genomes is often used in phylogenomic studies to measure the evolutionary distance between the genomes.

These four types of rearrangements can be modeled by $2$-breaks~\cite{alekseyev2008} (also called DCJs~\cite{yancopoulos2005}),
which break a genome at two positions and glue the resulting fragments in a new order.
They simplify the analysis of genome rearrangements and allow one to efficiently compute the corresponding evolutionary distance between two genomes.

\emph{Transpositions} represent yet another type of genome rearrangements that cuts off continuous segments of a genome and moves them to different positions.
In contrast to reversal-like rearrangements, transpositions are rarely observed and believed to appear in a small proportion in the course of evolution 
(e.g., in Drosophila evolution transpositions are estimated to constitute less than $10\%$ of genome rearrangements~\cite{ranz2003rearrangement}). 
Furthermore, transpositions are hard to analyze; in particular, computing the transposition distance is known to be NP-complete~\cite{bulteau2012}.
To simplify analysis of transpositions, they can be modeled by $3$-breaks~\cite{alekseyev2008} that break the genome 
at \textit{three} positions and glue the resulting fragments in a new order.

In the current work we propose a computational method for determining the proportion of transpositions (modeled as 3-breaks)
among the genome rearrangements (2-breaks and 3-breaks) between two genomes.
To the best of our knowledge, previously the proportion of transpositions was studied only from the perspective 
of its bounding with the weighted distance model~ \cite{bader2007,fertin2009}, where reversal-like and transposition-like rearrangements are assigned different weights.
However, it was empirically observed~\cite{Blanchette1996} and then proved that the weighted distance model does not, in fact, achieve its design goal~\cite{jiang2011}.
We further remark that any approach to the analysis of genome rearrangements that controls the proportion of transpositions 
would need to rely on a biologically realistic value, which can be estimated with our method.

We applied our method for different pairs among the rat, macaque, and human genomes and estimated the transpositions rate in all pairs to be around 0.26.


\section{Background}
\begin{figure}%
\begin{center}
\includegraphics[width=\textwidth]{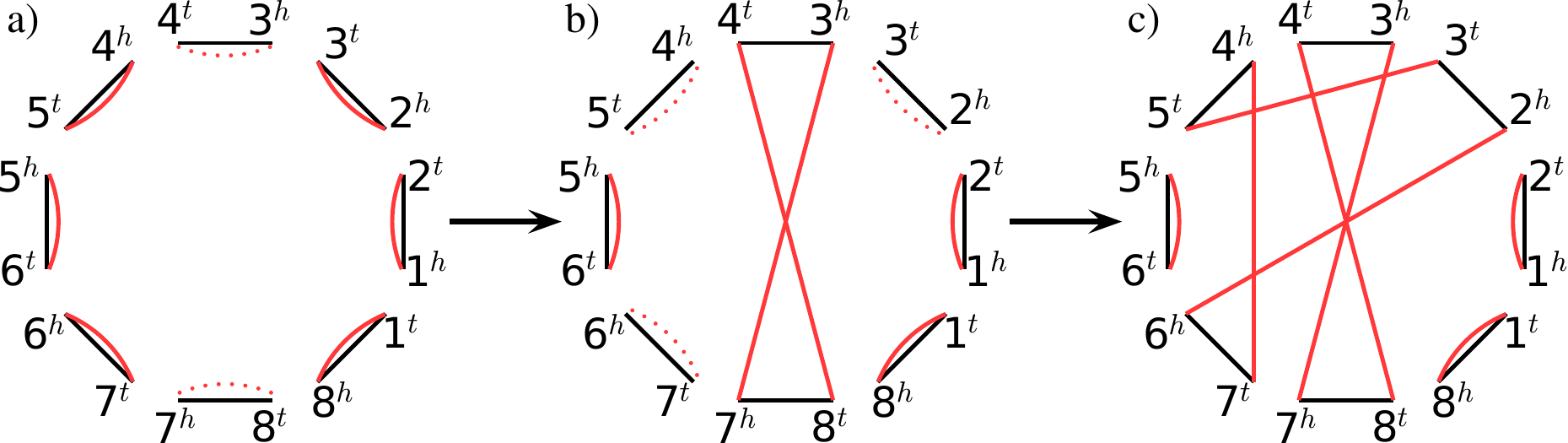}
 \caption{\textbf{a)} The breakpoint graph $G(P,Q_0)$ of ``black'' genome $P$ and ``red'' genome $Q_0=P$ , each consisting of a single circular chromosome $(1,2,3,4,5,6,7,8)$.
 Here, $n = 8$, $b = 0$, and all cycles in $G(P,Q_0)$ are trivial.
 \textbf{b)} The breakpoint graph of ``black'' genome $P$ and ``red'' genome $Q_1=(1,2,3,-7,-6,-5,-4,8)$ obtained from $Q_0$ with a reversal of a segment $4,5,6,7$ 
 (represented as 2-break on the dotted edges shown in a). Here we use $-i$ to 
 denote opposite orientation of the block $i$.
 The graph consists of $c_1 = 6$ trivial cycles and $c_2 = 1$ 2-cycle, and thus  $b =2c_2 = 2$.
 \textbf{c)} The breakpoint graph of ``black'' genome $P$ and ``red'' genome $Q_2=(1,2,-6,-5,3,-7,-4,8)$ obtained from $Q_1$ with a transposition of a segment $3,-7$  (represented as a single 3-break on the dotted edges shown in b).
 The graph consists of $c_1 = 3$ trivial cycles, $c_2 = 1$ 2-cycle, and $c_3 = 1$ 3-cycle; thus $b =  2c_2+3c_3 = 5$.}
 \label{fig:bpg}
 \end{center}
\end{figure}

For the sake of simplicity, we restrict our attention to circular genomes. 
We represent a genome with $n$ blocks as a graph which contains $n$ directed 
edges encoding blocks and $n$ undirected edges encoding block adjacencies. We denote the 
tail and head of a block $i$ by $i^t$ and $i^h$, respectively.
A \emph{2-break} replaces any pair of adjacency edges $\{x,y\}$, $\{u,v\}$ in the genome graph
with either a pair of edges $\{x,u\}$, $\{y,v\}$ or a pair of edges 
$\{u,y\}$, $\{v,x\}$. 
Similarly, a \emph{3-break} replaces any triple of adjacency edges with another triple of edges forming a matching on the same six vertices (Fig.~\ref{fig:bpg}).

Let $P$ and $Q$ be genomes on the same set $S$ of blocks (e.g., synteny blocks or orthologous genes). We assume 
that in their genome graphs the adjacency edges of $P$ are colored black and the adjacency edges of $Q$ 
are colored red.
The \emph{breakpoint graph} $G(P, Q)$ is defined on the set of vertices $\{i^t,i^h| i \in S\}$ 
with black and red edges inherited from genome graphs of $P$ and $Q$. 
The black and red edges in $G(P,Q)$ form a collection of alternating black-red cycles (Fig.~\ref{fig:bpg}). 
We say that a black-red cycle is an \emph{$\ell$-cycle} if it contains $\ell$ black edges (and $\ell$ red 
edges), and we denote the number of $\ell$-cycles in $G(P,Q)$ by $c_{\ell}(P,Q)$. 
We call $1$-cycles \textit{trivial cycles}\footnote{In the breakpoint graph constructed on synteny blocks, there are no trivial cycles since no adjacency 
is shared by both genomes. However, in our simulations below this condition may not hold, which would result in the appearance of trivial cycles.}
and we call \textit{breakpoints} the vertices belonging to non-trivial cycles.

The $2$-break distance between genomes $P$ and $Q$ is the minimum number of $2$-breaks required to transform $P$ into $Q$.
\begin{theorem}[\cite{yancopoulos2005}] 
The 2-break distance between circular genomes $P$ and $Q$ is 
$$d(P, Q) = n(P, Q) - c(P, Q) \; ,$$
where $n(P, Q )$ and $c(P, Q)$ are, respectively, the number of blocks and cycles in $G(P, Q)$.
\end{theorem}

While 2-breaks can be viewed as particular cases of 3-breaks (that keep one of the affected edges intact), 
from now on we will assume that 3-breaks change all three edges on which they operate.


\section{Estimation for the Transposition Rate}

In our model, we assume that the evolution represents a discrete Markov process, where
different types of genome rearrangements (2-breaks and 3-breaks) occur independently with fixed probabilities.
Let $p$ and $1-p$ be the rate (probability) of 3-breaks and 2-breaks, respectively.
For any two given genomes resulted from this process, our method estimates the value of $p$ as explained below.
In the next section we evaluate the accuracy of the proposed method on simulated genomes and further apply it to real mammalian genomes
to recover the proportion of transpositions in mammalian evolution.

Let the evolution process start from a ``black'' genome $P$ and result in a ``red'' genome $Q$. 
It can be viewed as a transformation of the breakpoint graph $G(P,P)$, where red edges are parallel to black edges and form trivial cycles,
into the breakpoint graph $G(P,Q)$ with 2-breaks and 3-breaks operating on red edges.
There are observable and hidden parameters of this process. Namely, we can observe the following parameters:
\begin{itemize}
 \item $c_{\ell}=c_{\ell}(P,Q)$, the number of $\ell$-cycles (for any $\ell \geq 2$) in $G(P,Q)$;
 \item $b=b(P,Q) = \sum_{\ell \geq 2} \ell c_{\ell}$, the number of active (broken) fragile regions between $P$ and $Q$, also equal the number of synteny blocks between $P$ and $Q$ and  the halved total length of all non-trivial cycles in $G(P,Q)$;
 \item $d=d(P,Q)$, the $2$-break distance between $P$ and $Q$;
\end{itemize}
while the hidden parameters are:
\begin{itemize}
 \item $n=n(P,Q)$, the number of (active and inactive) fragile regions in $P$ (or $Q$), 
 also equal the number of solid regions (blocks) and the halved total length of all cycles in $G(P,Q)$;
 \item $k_2$, the number of $2$-breaks between $P$ and $Q$,
 \item $k_3$, the number of $3$-breaks between $P$ and $Q$.
\end{itemize}
We estimate the rearrangement distance between genomes $P$ and $Q$ as $k_2+k_3$ and the rate $p$ of transpositions as
$$p = \frac{k_3}{k_2+k_3} \; .$$

We remark that in contrast to other probabilistic methods for estimation of evolutionary parameters (such as the evolutionary distance in~\cite{Lin08}),
in our method we assume that the number of trivial cycles $c_1$ is not observable. 
While trivial cycles can be observed in the breakpoint graph constructed on homologous gene families (rather than synteny blocks), 
their interpretation as conserved gene adjacencies (which happen to survive just by chance) implicitly adopts 
the \emph{random breakage model} (RBM)~\cite{Ohno70,Nadeau84} postulating that every adjacency has equal probability to be broken 
by rearrangements. The RBM however was recently refuted with the more accurate \emph{fragile breakage model} (FBM)~\cite{PevznerTeslerPNAS03} 
and then the \emph{turnover fragile breakable model} (TFBM)~\cite{Alekseyev10b}, 
which postulate that only certain (``fragile'') genomic regions are prone to genome rearrangements. 
The FBM is now supported by many studies (see \cite{Alekseyev10b} for further references and discussion).

\section{Estimation for the Hidden Parameters}

In this section, we estimate hidden parameters $n$, $k_2$, and $k_3$ using observable parameters, particularly $c_2$ and $c_3$.

Firstly, we find the probability that a red edge was never broken in the course of evolution between $P$ and $Q$. 
An edge is not broken by a single 2-break with the probability $\left(1-\frac{2}{n}\right)$ and by a single 3-break with the probability 
$\left(1-\frac{3}{n}\right)$. So, the probability for an edge to remain intact during the whole process of $k_2$ 2-breaks and $k_3$ 3-breaks is
$$ \left(1-\frac{2}{n}\right)^{k_2}\left(1-\frac{3}{n}\right)^{k_3} \approx e^{-\gamma} \; ,$$
where $\gamma = \frac{2k_2+3k_3}{n}$.

Secondly, we remark that for any fixed $\ell$, the number of $\ell$-cycles resulting from occasional splitting of longer cycles is negligible,\footnote{We remark
that under the parsimony condition long cycles are never split into smaller ones. Our method does not rely on the parsimony condition 
and can cope with such splits when their number is significantly smaller than the number of blocks.}
since the probability of such splitting has order $\frac{b}{n^2}$.
In particular, this implies that the number of trivial cycles (i.e., 1-cycles) in $G(P,Q)$ is approximately equal to the number of 
red edges that were never broken in the course of evolution between $P$ and $Q$. 
Since the probability of each red edge to remain intact is approximately $e^{-\gamma}$, the number of such edges is approximated by $n\cdot e^{-\gamma}$. 
On the other hand, the number of trivial cycles in $G(P,Q)$ is simply equal to $n-b$, the number of shared block adjacencies between $P$ and $Q$. That is,
\begin{equation}\label{eq:nb}
n-b \approx n  e^{-\gamma} \; .
\end{equation}

Thirdly, we estimate the number of 2-cycles in $G(P,Q)$. By the same reasoning as above, such cycles mostly result from 2-breaks that merge pairs of trivial cycles. 
The probability for a red edge to be involved in exactly one 2-break is
$\frac{2k_2}{n} \left(1-\frac{1}{n}\right)^{2k_2+3k_3-1}$. The probability that another red edge was involved in the same 2-break is
$\frac{1}{n} \left(1-\frac{1}{n}\right)^{2k_2+3k_3-1}$. Since the total number of edge pairs is $n(n-1)/2$, we have 
the following approximate equality for the number of 2-cycles: 
\begin{equation}\label{eq:c2}
c_2 \approx k_2  e^{-2\gamma} \; .
\end{equation}

And lastly,  we estimate the number of 3-cycles in $G(P,Q)$. As above, they mostly result from either 3-breaks that merge three 1-cycles, 
or 2-breaks that merge a 1-cycle and a 2-cycle. 
The number of 3-cycles of the former type approximately equals $k_3  e^{-3\gamma}$ analogously to the reasoning above. 
The number of 3-cycles of the latter type is estimated as follows. 
Clearly, one of the red edges in such a 3-cycle results from two 2-breaks, say $\rho_1$ followed by $\rho_2$, which happens with the probability about
$$\frac{2k_2(2k_2-2)}{2n^2}\left(1-\frac{1}{n}\right)^{2k_2+3k_3-2} \approx 2\frac{k^2_2}{n^2} e^{-\gamma} \; .$$
One of the other two edges results solely from $\rho_1$, while the remaining one results solely from $\rho_2$, 
which happens with the probability about $\left(\frac{1}{n}e^{-\gamma}\right)^2$. 
Since there are about $n^3$ ordered triples of edges, we get the following approximate equality for the number of 3-cycles: 
\begin{equation}\label{eq:c3}
c_3\approx k_3 e^{-3\gamma}+\frac{2k_2^2}{n}e^{-3\gamma} \; .
\end{equation}

\begin{figure}%
\begin{center}
 \includegraphics[width=\textwidth]{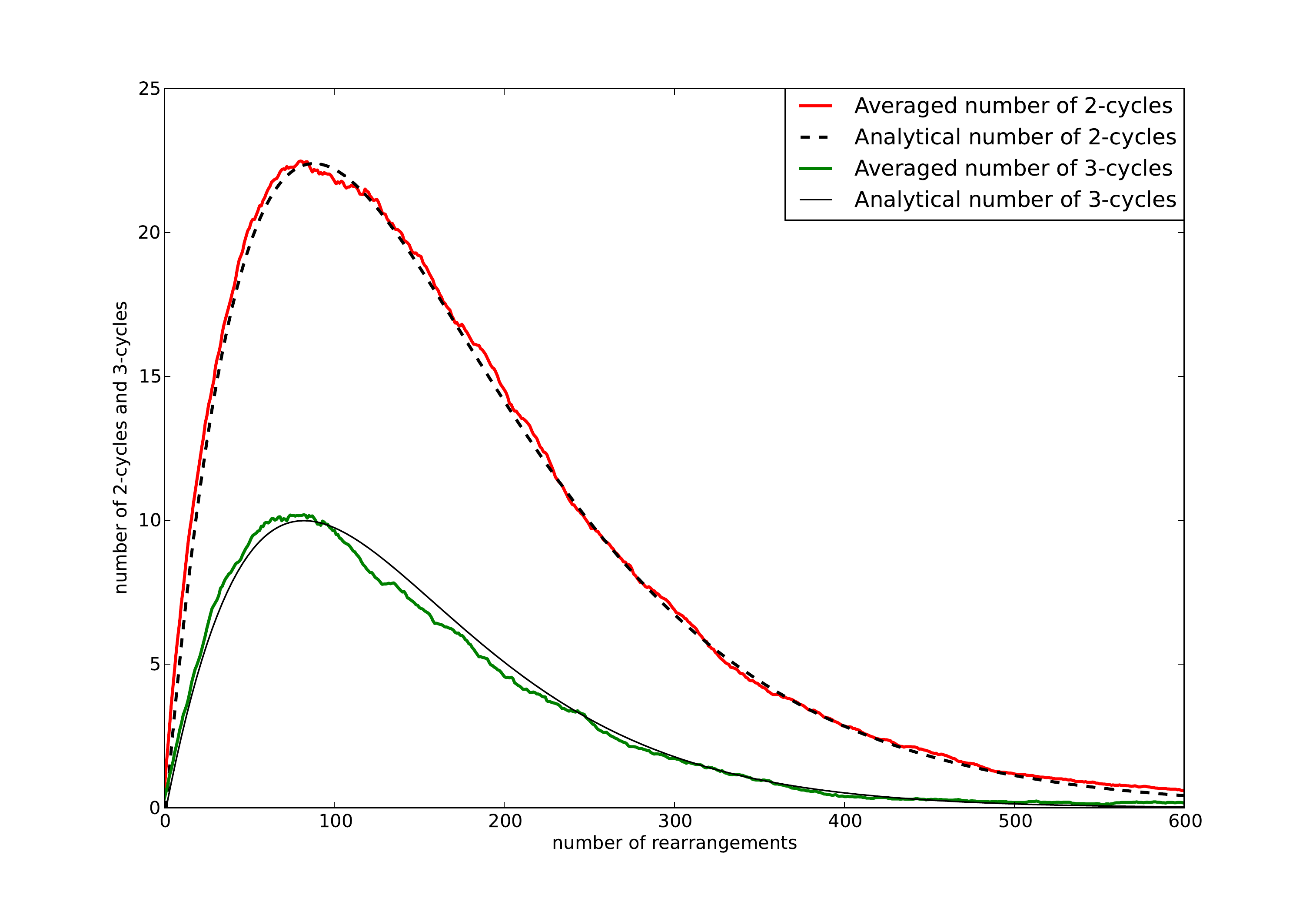}
 \caption{Empirical and analytical curves for the number of 2-cycles and 3-cycles averaged over 100 simulations on $n=400$ blocks 
 with proportion of 3-breaks $p=0.3$.}
 \label{fig:c2c3}
 \end{center}
\end{figure}

Fig.~\ref{fig:c2c3} provides an empirical evaluation of the estimates (\ref{eq:c2}) and (\ref{eq:c3}) for the number of 2-cycles and 3-cycles in $G(P,Q)$, 
which demonstrates that these estimates are quite accurate.

Below we show how one can estimate the probability $p$ from the (approximate) equations \eqref{eq:nb}, \eqref{eq:c2}, and \eqref{eq:c3}.

We eliminate $k_2$ from \eqref{eq:c3}, using \eqref{eq:c2}:
$$ c_3 \approx k_3  e^{-3\gamma}+\frac{2c_2^2}{n} e^{\gamma} \; .$$
Now we consider the following linear combination of the last equation and \eqref{eq:c2}:
$$2e^{-\gamma}(c_2 - k_2 e^{-2\gamma})+3\left(c_3 -(k_3  e^{-3\gamma}+\frac{2c_2^2}{n} e^{\gamma})\right)\approx 0 \; .$$
It gives us the following equation for $\gamma$ and $n$:
$$\gamma e^{-3\gamma} \approx \frac{1}{n}\left(2c_2 e^{-\gamma}+3c_3-\frac{6c_2^2e^\gamma}{n}\right) \; .$$
Using \eqref{eq:nb}, we eliminate $n$ from the last equation and obtain the following equation with respect to a single indeterminate $\gamma$:
\begin{equation}\label{eq:gamma}
\gamma e^{-3\gamma} \approx \frac{1-e^{-\gamma}}{b}\left(2c_2 e^{-\gamma}+3c_3-\frac{6c_2^2e^\gamma(1-e^{-\gamma})}{b}\right) \; .
\end{equation}


Solving this equation numerically (see Example \ref{example}, Section \ref{sec:simulated}), we obtain the numerical values 
for $\gamma^{est}$, $n^{est}$, $k^{est}_2$ and $k^{est}_3$, and, finally, 
$$p_{est} = \frac{k_3^{est}}{k^{est}_2+k^{est}_3} \; .$$

\section{Experiments and Evaluation}

\subsection{Simulated Genomes} \label{sec:simulated}

\begin{figure}%
\begin{center}
 \includegraphics[width=0.8\textwidth]{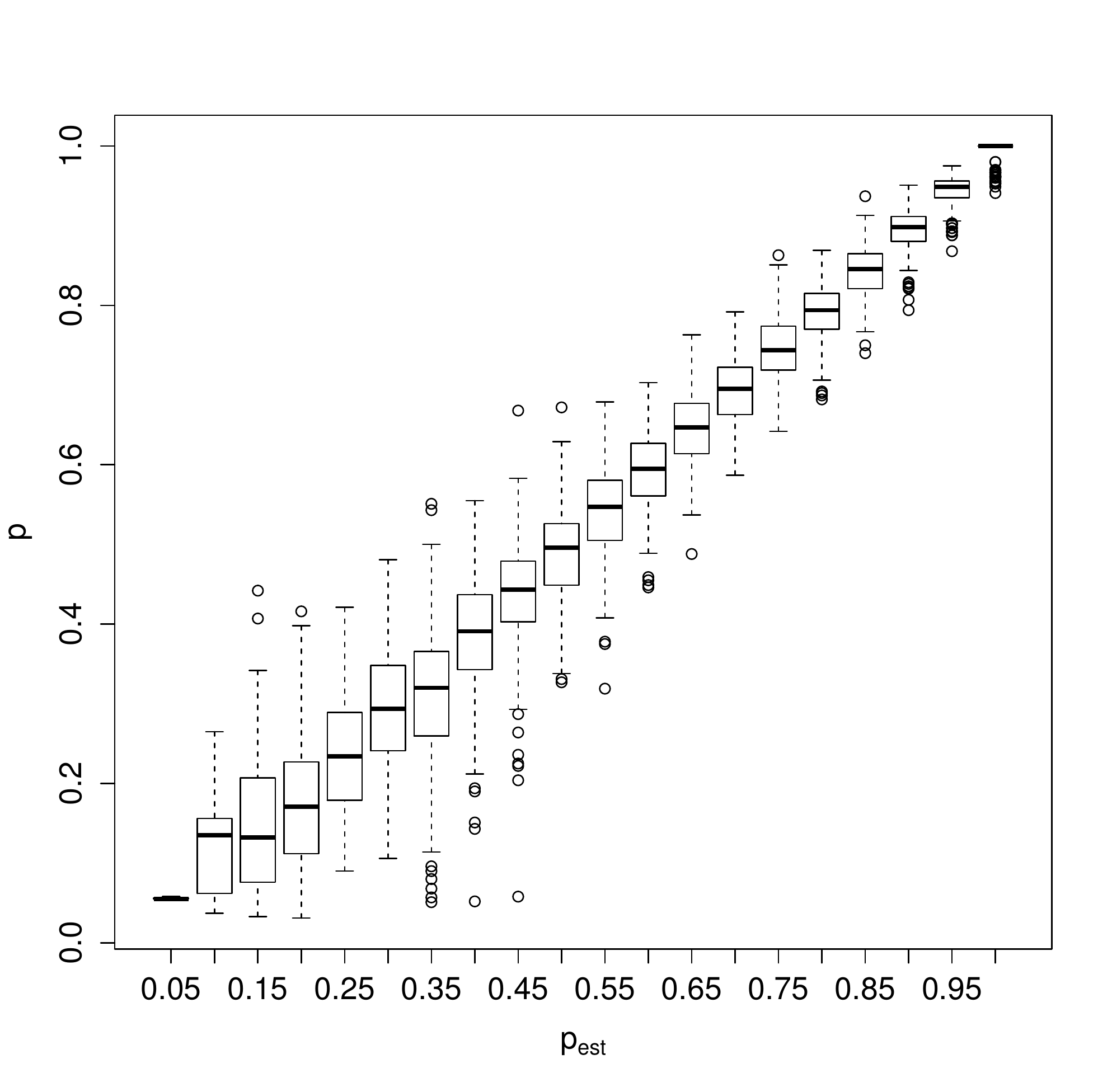}
 \caption{Boxplots for the value of $p$ as a function of $p_{est}$}
 \label{fig:estp}
\end{center}
\end{figure}

We performed a simulation with a fixed number of blocks $n = 1800$ 
and variable parameters $p$ and $\gamma$.
In each simulation, we started with a genome $P$ and applied a number of 2-breaks and 3-breaks with probability $1-p$ and $p$, respectively, until 
we reached the chosen value of $\gamma$. We denote the resulting genome by $Q$ and estimate $p$ with our method as $p_{est}$.
We observed that the robustness of our method mostly depends on $p$ and $\gamma$, and it becomes unstable for $p_{est}<0.15$ (Fig.~\ref{fig:estp}).
So in our experiments we let $p$ range between $0.05$ and $1$ with step $0.05$ and $\gamma$ range between $0.2$ and $1.2$ with step $0.1$.

In Fig.~\ref{fig:estp}, we present boxplots for the value of $p$ as a function of $p_{est}$ cumulative over the values of $\gamma$.
These evaluations demonstrate that $p_{est}$ estimates $p$ quite accurately with the absolute error below $0.1$ in $90\%$ of observations.

\begin{example}
\label{example}
Let us consider the example from our simulated dataset. In this example, the number of active blocks 
$b = 716$, the number of 2-cycles $c_2 = 107$, the number of 3-cycles $c_3 = 48$, and the hidden parameters are: 
 the total number of blocks is $n = 1800$,  the number of $2$-breaks $k_2 = 279$ and the number of $3$-breaks is $k_3 = 114$.
 So, the value of $p$ in this example is $0.29$ and the value of $\gamma$ is 0.5.
 
\begin{figure}
  \begin{center}
   \includegraphics[width=0.8\textwidth]{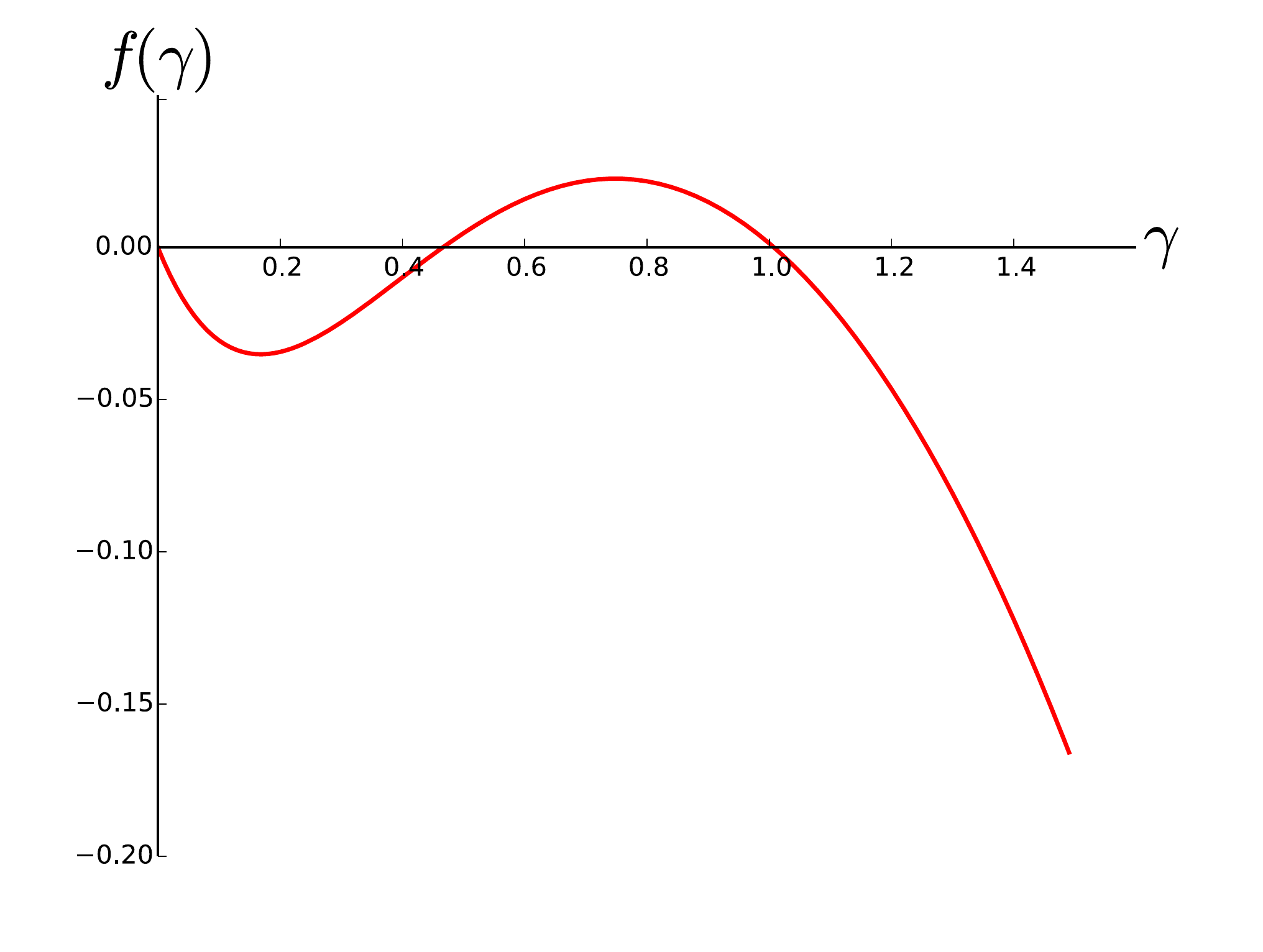}
    \caption{Typical behavior of
    $f(\gamma) = \gamma e^{-3\gamma}-\frac{1-e^{-\gamma}}{b}\left(2c_2 e^{-\gamma}+3c_3-\frac{6c_2^2e^\gamma(1-e^{-\gamma})}{b}\right)$, 
    the difference between right and left hand sides of \eqref{eq:gamma}, where $b = 716$, $c_2 = 107$, $c_3 = 48$.}
    \label{fig:gamma}
    \end{center}
\end{figure}
At first, using the bisection method, one can find roots of \eqref{eq:gamma}. In this case there are two roots:
$\gamma = 0.466$ and $\gamma = 1.007$ (See Fig.\ref{fig:gamma}). Let us check the root $0.466$ first.
Then, using \eqref{eq:nb}, one finds the estimated value of $n$: $716/(1-e^{0.466})\approx 1922$. Equation \eqref{eq:c2} gives us the  estimated value of $k_2$: $107 e^{2\cdot0.466} \approx 272$.
One can estimate $k_3$ as $(\gamma n - 2k_2)/3 \approx 117$.
And finally we obtain the estimated value of $p$: $117/(117+272) \approx 0.3$.

In this example, using the second root of \eqref{eq:gamma} yields a negative value for $k_3$, so we do not consider it.
So, our method quite accurately estimates the value of $p$, and also values of $\gamma$ and $n$.
\end{example}

\subsection{Mammalian Genomes}

We analyzed a set of three mammalian genomes: rat, macaque, and human, represented 
as sequences of $1,360$ synteny blocks~\cite{ma2006,alekseyev2009}.
For each pair of genomes, we circularized\footnote{While chromosome circularization introduces artificial edges to the breakpoint graph, 
the number of such edges (equal to the number of chromosomes) is negligible as compared to the number of edges representing block adjacencies in the genomes. For subtle differences in analysis of circular and linear genomes see \cite{alekseyev2008multi}} 
their chromosomes, constructed the breakpoint graph, obtained parameters $b$, $c_2$, $c_3$, 
and independently estimated the value of $p$.
The results in Table~\ref{tab:mam} demonstrate consistency and robustness with respect to the evolutionary distance between the genomes 
(e.g., the 2-break distance between rat and human genomes is 714, while the 2-break distance between macaque and human genomes is 106). 
The rate of transpositions for all genome pairs is estimated to be around 0.26.
Numerical experiments suggest that the 95\% confidence interval for such values is [0.1, 0,4] (Fig.~\ref{fig:estp}).

\begin{table}
\caption{Observable parameters $b$, $c_2$, $c_3$ and estimation $p_{est}$ for the rate of evolutionary transpositions 
between circularized rat, macaque, and human genomes.}
\label{tab:mam} 
\centering
\begin{tabular}{| c || c | c | c | c | }
\hline
  Genome pair & $b$ &   $c_2$ &    $c_3$ & $p_{est}$ \\
  \hline\hline
  rat-macaque & 1014 & 201 & 85 & 0.27 \\
  \hline
  rat-human & 1009 & 194 & 79 & 0.26 \\
  \hline
  macaque-human & 175 & 45 & 17 & 0.25 \\
  \hline
\end{tabular}

\end{table}

\section{Discussion}
In the present work we describe a first computational method for estimation of the transposition rate
between two genomes from the distribution of cycle lengths in their breakpoint graph. 
Our method is based on modeling the evolution as a Markov process under the assumption that the transposition rate remains constant.
The method does rely on the random breakage model \cite{Ohno70,Nadeau84} and thus is consistent with more prominent fragile breakage model \cite{PevznerTeslerPNAS03,Alekseyev10b} of chromosome evolution.
As a by-product, the method can also estimate the true rearrangement distance (as $k_2+k_3$) in the evolutionary model that includes 
both reversal-like and transposition-like operations.

Application of our method on different pairs of mammalian genomes reveals that the transposition rate is almost the same 
for distant genomes (such as rat and human genomes)
and close genomes (such as macaque and human genomes), suggesting that the transposition rate remains the same across different lineages in mammalian evolution.

In further development of our method, we plan to employ the technique of stochastic differential equations, 
which may lead to a more comprehensive description of the $c_{\ell}$ behavior.
It appears to be possible to obtain equations, analogous to \eqref{eq:c2} and \eqref{eq:c3}, 
for $c_\ell$ with $\ell>3$. This could allow one to verify the model and estimate the transposition rate more accurately.

\subsection*{Acknowledgements}
The authors thank Jian Ma for providing the synteny blocks for mammalian genomes.

The work is supported by the National Science Foundation under the grant No. IIS-1462107. The work of
NA is also supported by grant 6.38.672.2013 of SPbSU and RFFB grant 13-01-12422-ofi-m.
 
\bibliographystyle{splncs}
\bibliography{alpha.bib}

\end{document}